\documentstyle[12pt,epsfig]{article}
\begin{document}
\title{A Simple Passive Scalar Advection-Diffusion Model}
\author{Scott Wunsch}
\maketitle

\centerline{{\it The James Franck Institute, The University of Chicago}}

\centerline{{\it 5640 South Ellis, Chicago, IL 60637, USA}}

\bigskip

\section{Abstract}
This paper presents a simple, one-dimensional model of a randomly
advected passive scalar.  The model exhibits anomalous inertial range 
scaling for the structure functions constructed from scalar differences.
The model provides a simple computational test for recent ideas regarding
closure and scaling for randomly advected passive scalars.  Results suggest
that high order structure function scaling depends on the largest velocity eddy
size, and hence scaling exponents may be geometry-dependent and
non-universal.

{\it PACSs:} 47.27.GS; 05.40.+j

{\it Keywords:} Isotropic turbulence; Random processes; Passive scalars

\section{Introduction}
Passive scalars are `tracer particles' which are advected by the flow of
another fluid.  A typical example might be ink mixed by flowing water.
The mixing of a scalar field $T$ in a velocity field ${\bf u}$ is governed by 
the advection-diffusion equation,
\begin{equation}
	\left( \partial_t +  {\bf u}({\bf x},t) \cdot \nabla
	\right)T({\bf x},t)  = D \nabla^2 T({\bf x},t) +f({\bf x},t)
\end{equation}
in which $D$ is a diffusion constant and $f$ is an external source of scalar.
The velocity field is determined independently of the scalar field (hence the
adjective {\it passive}) and (in this discussion) is incompressible (${\bf\nabla \cdot u}=0$).
The incompressible flow simply transports the scalar, without changing any
of the scalar's values.  The diffusion term smooths the resulting scalar 
differences, while the source term generates new scalar to compensate for this
smoothing.   

The mixing of a scalar field $T$ is characterized experimentally by its
structure functions, defined as $S_{2n} ({\bf r})  \equiv \langle 
(T({\bf x+r},t)-T({ \bf x},t))^{2n} \rangle$ for positive integers $n$.
It has been established experimentally that passive scalar structure
functions are scaling functions of $r$ over a wide range of
length scales when the advecting flow is turbulent.
This range of length scales, known as the inertial range, 
extends from a macroscale (usually determined by the flow geometry)
down to a dissipative cut-off scale.  The lower end of the inertial
scaling range is set either by the viscous scale of the velocity 
field (the scale where viscosity damps out the fluid motion) or by
the scalar's dissipation scale (the scale where molecular diffusivity
damps out scalar fluctuations).  The ratio of these length scales, known as
the Prandtl number, is a material property of the system.  This 
work will focus exclusively on systems in which the molecular diffusivity
imposes the lower bound on the inertial range (Prandtl number less than one).
In such systems the size of the inertial range is measured by the Peclet
number, defined as the ratio of the macroscale to the dissipation scale. 

The inertial range scaling exponents $\rho_{2n}$, defined by 
$S_{2n} (r) \sim r^{\rho_{2n}}$, are measured experimentally.
The simplest possible prediction is based on the 
belief that the scalar difference between two points separated by a distance
$r$ will typically have a magnitude $\Delta T(r)$ which scales with 
$r$, $\Delta T(r) \sim r^\beta$ (for some $\beta$).  This implies 
$S_{2n} (r) \sim r^{2n\beta}$.  Consequently, the scaling exponents
satisfy $\rho_{2n} = n\rho_2$, which is known as `regular scaling.'
Experimentally, the exponents are seen to exhibit `anomalous scaling,'
$\rho_{2n} < n\rho_2$.  Such scaling indicates that there is no                      
typical magnitude for $\Delta T(r)$, and this wide variation in
magnitudes is known as {\it multiscaling}.  Experimentally, measured scaling 
indices for passive scalars in turbulent flows are quite anomalous. \cite{Ant84}
Until recently, relatively little attention was paid                       
to this observed anomalous scaling in the passive scalar field.  This                              
is perhaps because the phenomena had often been attributed to anomalous                            
scaling of the underlying velocity field.    

However, a model proposed by Robert Kraichnan \cite{kra94} 
suggests that a hypothetical regular-scaling velocity field
would still generate anomalous scaling in the scalar structure
functions.  In addition, the scaling exponents would be universal,
independent of details such as the geometry of the flow. 
In this model, the passive scalar is advected by a stochastic
velocity field with regular scaling in space but with an
extremely short correlation time.  It is possible to form equations 
for the structure functions in this model, provided a closure 
assumption is made.  The proposed closure \cite{kra94} \cite{kra95} 
leads to structure functions which scale with $r$, but
with scaling exponents $\rho_{2n}$ which grow like $\sqrt{n}$ rather
than $n$.  Others, however, have applied different techniques to 
the same model (avoiding a closure ansatz) and claim to have found
different results. \cite{gaw95} \cite{shr95} \cite{che95} \cite{che96}
A review of the disagreement is given by Shraiman and Siggia. \cite{shr96}
   
To study this closure ansatz and the passive scalar structure functions
numerically, I use a one-dimensional model \cite{kad97}
in which the structure functions
$S_{2n}$ obey statistical equations with the same closure problem as
in Kraichnan's model.  The model is simple enough to allow direct
numerical study of the structure function scaling (over 2 orders of 
magnitude in the separation $r$) and the proposed closure ansatz.
Mixing occurs through a random mapping function chosen
to mimic an incompressible flow and induce scaling
in the structure functions.  Because it is restricted to one
dimension, the model is best viewed as representing scalar mixing in
a turbulent pipe flow.  
    
In this model the second order structure function can be determined
analytically.  For large Peclet
number it scales with $r$ with a scaling exponent $\rho_2$
determined by the advective mapping function.  The higher 
order structure functions $S_{2n}(r)$ also scale (numerically),
with exponents $\rho_{2n}$ which appear to approach a constant
value at large $n$. 

The closure ansatz proposed by Kraichnan is seen to fail in numerical
simulations of this model.  The source of this failure can 
be traced to the finite size of the largest mixing events.
This upper length scale is analogous to the largest eddy size in
a fluid flow (in a pipe this size is determined by the pipe diameter).  
Very large scalar differences 
cannot be produced by a single mixing event, but instead arise from
the combined action of several events.  For this reason
large scalar differences are exponentially improbable.  Deviations
from the closure ansatz are observed only for these large scalar
differences.  However, the behavior of the large $n$ structure 
functions is determined by these rare events.  In this model
the scaling exponents approach a constant at large $n$, in contrast
to the $\sqrt n $ behavior predicted for Kraichnan's model.  In addition,
the value of the constant might depend on the largest eddy size, so that
the large $n$ behavior of the exponents would be geometry dependent. 

Because these conclusions rest primarily on the fact that the largest mixing
events are much smaller than the system size, one might expect them to be
independent of the details of the small-scale mixing process.  Mixing in
a turbulent pipe flow should exhibit this separation of length scales, and
there is some experimental evidence to support the idea that large scalar 
differences are exponentially unlikely in such a flow. \cite{Gui97}

\subsection{Kraichnan's Passive Scalar Model}

Robert Kraichnan first introduced the `white-advected' passive scalar
model in 1968 \cite{kra68},  and introduced the anomalous-scaling
solution in a pair of recent papers. \cite{kra94} \cite{kra95}  All 
of the principal results of this section can be found in the recent
literature, although some of the notation has been altered here.

In this model the passive scalar $T({\bf x},t)$ obeys the usual 
advection-diffusion equation, 
\begin{equation}
	\left( \partial_t + {\bf u}({\bf x},t) \cdot \nabla
	\right)T({\bf x},t)  = D \nabla^2 T({\bf x},t) +f({\bf x},t)
\end{equation}
in which $D$ is a diffusion constant.  The external source $f$ was not 
explicitly considered by Kraichnan, but is necessary to maintain a state of
statistical equilibrium.  The incompressible velocity field 
(${\bf\nabla \cdot u}=0$) is random with zero mean and is white in time.
The second order velocity structure function is specified as
\begin{equation}
	\langle [{\bf u}({\bf x+r},t)-{\bf u}({\bf x},t)] \cdot 
	{{\bf r} \over r} [{\bf u}({\bf x+r},t')-{\bf u}({\bf x},t')]
	\cdot {{\bf r} \over r} \rangle \equiv C \delta(t-t')
	\left( {r \over L_v} \right) ^\eta
\end{equation}
where $C$ is a constant and $\eta$ is a parameter of the model.  This
scaling in space holds until $r$ approaches some large correlation length
$L_v$, representing the largest motions in the system.
The scaling correlation in space is similar to real turbulent flows, while 
the very short range of temporal correlations is unphysical but desirable for
technical reasons. The velocity statistics are chosen to be gaussian, so higher 
order velocity structure functions can be expressed in terms of this second 
order function.  Consequently, the velocity structure functions exhibit
regular scaling, and so any anomalous scaling found in the passive scalar
structure functions must arise from the structure of the scalar
advection-diffusion equation and not from the underlying velocity scaling.

The source term is also a gaussian random variable with zero mean.
Its correlations are specified by
\begin{equation}
	\langle f({\bf x+r},t)f({\bf x},t') \rangle
	\equiv \chi (r) \delta (t-t').
\end{equation}
The spatial function $\chi (r)$ is assumed to be smooth, so that 
$\chi (r) \simeq \chi (0)$ for $r \ll L_s$, where $L_s$ represents the
correlation length of the source field.  

Calculation of the structure functions proceeds by forming an equation
for scalar differences at equal times, defined as
$\Delta ({\bf x,y};t) \equiv T({\bf x},t) - T({\bf y},t)$.
One can take the ensemble average of this equation by integrating over
the recent history of the scalar and using the defined correlation functions
for $\bf u$ and $f$.  The $\delta$-correlation in time makes it possible
to do the resulting integrals.  Using the definition 
$S_{2n}({\bf x-y}) \equiv \langle \Delta ^{2n} ({\bf x,y}) \rangle$, the 
structure functions are found to obey
\begin{equation}\label{StatEq}
	\left( \partial_t + {\cal L} \right) S_{2n} = J_{2n} + F_{2n}. 
\end{equation}
In this equation, $\cal L$ is the Richardson eddy-diffusivity operator, 
defined as 
\begin{equation}
	{\cal L} S_{2n}(r) \equiv -2C L_v^{-\eta} r^{1-d} 
	\partial_r (r^{\eta + d-1} \partial_r S_{2n}(r))
\end{equation}
where $r \equiv |{\bf x-y}|$ and $d$ is the dimension of space.  
The source term
\begin{equation}
	F_{2n} \equiv 2n(2n-1)S_{2n-2}(\chi (0) - \chi (r)) 
\end{equation}
was makes it possible to establish a state of statistical equilibrium
($\partial_t S_{2n}=0$).  The dissipation function
\begin{equation} \label{JDef}
	J_{2n} \equiv 2nD \langle \Delta^{2n-1} (\nabla_x^2 + \nabla_y^2)
	\Delta \rangle 
\end{equation}
must be expressed in terms of $S_{2n}$ before the system of equations for 
$S_{2n}$ will be closed. 

The lowest non-trivial structure function, $n=1$, can be determined exactly.
The dissipation function $J_2$ can be evaluated by commuting derivatives with
$\Delta$ and with the averaging process $\langle ... \rangle$, so that
\begin{equation}
	J_2 = 2D\nabla_r^2 S_2(r) - 4D\langle (\nabla T)^2 \rangle .
\end{equation}
The mean-squared scalar dissipation $D\langle (\nabla T)^2 \rangle$ is constant, 
and can be determined by balancing against the scalar source (since the flux
of scalar is conserved).  At very large separations
$r \gg L_s$, the source correlation vanishes ($\chi(r) \to 0$) and
$S_2$ must approach a constant.  Consequently, the equation for $S_2$ implies
$2D\langle (\nabla T)^2 \rangle = \chi (0)$.
The complete equation for $S_2$ can then be re-cast as
\begin{equation}
	Cr^{1-d}\partial_r (r^{d-1} \left( {r \over L_v} \right) ^\eta 
	\partial_r S_{2}(r)) + Dr^{1-d}\partial_r (r^{d-1} \partial_r S_{2}(r)) = \chi (0)
\end{equation}
assumming $\chi (r) \simeq \chi (0)$. 
The two differential operators which act on $S_2$ possess different scalings.
The length scale at which the operators are equal is defined as the dissipation
scale, $r_d$:
\begin{equation}
	r_d \equiv L_v \left( {D \over C} \right) ^{1 \over \eta}. 
\end{equation}
The upper limit of the inertial (scaling) range is set by the lesser of the
two macroscales, $L_s$ and $L_v$.  In the inertial range ($r_d  \ll r \ll min(L_s,L_v)$), 
the second operator can be neglected, and the solution is approximately
\begin{equation}
	S_2(r) \simeq {\chi (0) L_v^2 \over C \rho_2 d} 
	\left( {r \over L_v} \right) ^{\rho_2}
\end{equation}
where $\rho_2 \equiv 2 - \eta$ is the second-order scaling exponent.
Alternately, in the dissipation range ($r \ll r_d$), the first operator
can be neglected, and the solution is approximately
\begin{equation}
	 S_2(r) \simeq {\chi (0) \over 2Dd} r^2 .
\end{equation}

There are two possible approaches for determining the higher order
scaling exponents.  The first is to balance the eddy-diffusivity
operator against the forcing term.  If the
$2n$-th structure function has scaling exponent $\rho_{2n}$, then 
the corresponding eddy-diffusivity term scales with $r$ with the
scaling exponent $\rho_{2n} - \rho_2$.  The source term has
scaling $\rho_{2n-2}$.  If these terms are equal ({\it i.e.} if 
$J_{2n}$ is negligible) then the scaling exponents are given by
$\rho_{2n} = n \rho_2$, which is regular scaling. 

The other possibility, first proposed by Kraichnan \cite{kra94}, is
to balance the eddy-diffusivity operator against the dissipation
term $J_{2n}$.  This basically assumes that the dominant scaling
is set not by the source term, but by a homogeneous solution (or 
{\it zero mode}) of the structure function equations.  Consequently, 
the scaling could be {\it universal}, independent of the details
of the source term.  

Unfortunately, evaluation of the $J_{2n}$ requires a closure ansatz stating
how the laplacian of the scalar field at two distinct points, 
$(\nabla_x^2 + \nabla_y^2) \Delta ({\bf x,y})$ depends
on the scalar difference $\Delta ({\bf x,y})$ between those points.
The proposed closure ansatz is best understood in terms of the probability 
distribution function (PDF) for scalar differences $P(\Delta,r)$.  This
function gives the probability for seeing a particular value $\Delta$
for the scalar difference between two points separated by a distance $r$. 
The structure functions $S_{2n}(r)$ are the moments of this PDF:
\begin{equation}
	S_{2n} (r) \equiv {\int P(\Delta,r) \Delta ^{2n} d\Delta}.
\end{equation}
The unknown closure information is expressed in the conditional probability
\begin{equation}
	 H(\Delta,{\bf x-y}) \equiv \langle (\nabla_x^2+ \nabla_y^2)
	 \Delta({\bf x,y})  \vert  \Delta({\bf x,y}) \rangle
\end{equation}
which gives the ensemble-averaged value of 
$(\nabla_x^2 + \nabla_y^2) \Delta ({\bf x,y})$
subject to the constraint that the scalar difference between
the two points take on the particular value $\Delta$.  The dissipation
functions can be expressed using $H$ and the PDF for scalar differences
$P(\Delta,r)$:
\begin{equation}
	J_{2n} (r)= 2nD{\int P(\Delta,r) \Delta ^{2n-1} H(\Delta,r)  d\Delta}.
\end{equation}
The closure ansatz proposed in \cite{kra95} is that $H$ is approximately
a linear function of $\Delta$:
\begin{equation}
	H(\Delta,r) \cong \alpha (r) \Delta
\end{equation}
with some unknown slope $\alpha$.  This assumption can be viewed as a 
truncation of the Taylor series expansion of $H(\Delta)$ (by symmetry, the
series has only odd terms).  By substitution, the dissipation
functions are $J_{2n} = 2nD \alpha S_{2n}$.  The function $\alpha$ can 
be determined by considering the $n=1$ case.  The result is that
\begin{equation} \label{J2n}
	J_{2n} = nJ_2 {S_{2n} \over S_2}
\end{equation}
(recall $J_2$ is approximately constant in the inertial range). 
Hence both $J_{2n}$ and ${\cal L}S_{2n}$ have scaling exponent
$\rho_{2n} - \rho_2$.  Consequently, the scaling exponents are
set by the {\it coefficients} of the terms.  The result is that
\begin{equation} \label{KraExp}
	\rho_{2n} = {1 \over 2} \sqrt{4nd\rho_2 + (d - \rho_2)^2}
	- {1 \over 2} (d - \rho_2).
\end{equation}
The asymptotic behavior is that $\rho_{2n} \simeq \sqrt{nd\rho_2}$, 
which depends only on the spatial dimension $d$ and the velocity 
scaling $\eta$ (through $\rho_2 = 2 - \eta$).  

\subsection{Competing Calculations}

The anomalous scaling for $\rho_{2n}$ in the Kraichnan model depends
on the closure ansatz, $H(\Delta) \propto \Delta$.  Unfortunately, it is 
not possible to consider higher order corrections to the Taylor series for
$H(\Delta)$ because the resulting equations for $S_{2n}$ cannot be solved.  
However, others have approached the model using techniques which avoid the
need for a closure ansatz and, in certain limits, have found values of 
$\rho_{2n}$ which disagree with Kraichnan's result.   

The competing approach to determining scaling exponents in this model
focuses on the $n$-point correlation functions of the scalar (known as 
$n$th order moments), defined as
\begin{equation}
	M_{n}({\bf x_1, x_2, ... x_n}) \equiv \langle T({\bf x_1})
	T({\bf x_2}) ... T({\bf x_n}) \rangle. 
\end{equation} 
The $n$th order structure function can be expressed in terms of the
corresponding moment by allowing points to fuse together.  The moments
convey more information than the structure functions and can be viewed
as the more fundamental objects, but moments are generally not measured
experimentally.  

The $\delta$-correlation in time makes it possible to construct closed
equations for the moments (known as {\it Hopf} equations).  There is no
closure problem associated with the dissipative terms.  Unfortunately, 
the equation of the $n$th order moment is a partial differential equation
in $n!$ vector separations, and it is beyond the reach of known mathematics
to determine the solutions of such complicated equations.  

To determine scaling exponents, several groups have assumed an overall
scaling for the $n$th order moment when all separations lie within the 
inertial range, and then attempted to calculate the scaling
exponent in certain limits using perturbation theory.  In each case it 
is assumed that the dominant
contribution comes from a zero mode of the Hopf equation.  These exponents
differ from the structure function scaling exponents proposed by Kraichnan.
The most readable description of this approach can be found in the paper
by Chertkov, {\it et. al.}.  \cite{che95}

One of the first of these calculations was done in the limit $\rho_2 \to 2$.
\cite{gaw95}  The result was that the 4th order scaling exponent was given by 
\begin{equation}
	\rho_4 \simeq 2\rho_2 - {4 \over d+2} \eta
\end{equation}
($\eta = 2-\rho_2 \ll 1$).  This 
indicates that the scaling is nearly regular in this limit, with a small
correction for $\rho_2 < 2$.  The solution proposed by Kraichnan (equation 
\ref{KraExp}) is not regular in this limit;  instead it yields
\begin{equation}
	\rho_4 = \left( {1 \over 2} \sqrt{d^2 + 12d + 4} + 1 - {1 \over 2}d \right)
	- {1 \over 2} \left( {2+3d \over \sqrt{d^2 + 12d +4}} -1 \right) \eta
\end{equation}
which is quite anomalous as $\eta \to 0$ (in three dimensions $\rho_4 \to 3$, not 4).
This disagreement is very substantial.  Unfortunately, the huge difference in 
approach makes it difficult to ascertain the source of the disagreement. 

Another limit in which a descrepancy was found is the limit of large space dimension
($d \to \infty$). \cite{che95} \cite{che96}  In this case the scaling is
again nearly regular, with a small correction in $1/d$.  The corresponding limit
of equation \ref{KraExp} is also regular, but the correction term differs from 
that obtained from the Hopf equation calculation.

One proposed reconcilliation between these calculations is a generalization
of Kraichnan's closure based on the `fusion rules.' \cite{chi96}
The dissipation functions are given by a generalized form of equation \ref{J2n}
\begin{equation}
	 J_{2n} = nC_nJ_2 {S_{2n} \over S_2}
\end{equation}
with an unknown constant of proportionality $C_n$.  Values $C_n \ne 1$ would
change the result for the scaling exponents to
\begin{equation} \label{ModExp}
	\rho_{2n} = {1 \over 2} \sqrt{4nC_nd\rho_2 + (d - \rho_2)^2}
        - {1 \over 2} (d - \rho_2).
\end{equation}
No precise form for $C_n$ has been proposed, but Ching \cite{chi97} argues that 
$C_n \propto n$ for large $n$, indicating $\rho_{2n} \propto n$
for large $n$. 

Numerical simulations of the Kraichnan model in two dimensions have been
attempted by two groups. \cite{kra95} \cite{fai97}  Although the size of 
the inertial range is limited, both groups claim to support Kraichnan's 
solution for intermediate values of $\rho_2$. The limit $\rho_2 \to 2$
has not been approached in either case.

\section{A Simple One-Dimensional Model}
\subsection{Definition of the Model}
To address the debate, I propose a simple passive scalar model
which reproduces the closure problem described above. The closure ansatz
can be studied numerically on two levels.  First, one can calculate
the dissipation functions $J_{2n}$ and measure the constants of proportionality
to determine if $C_n \ne 1$.  Second, one can study the conditional probability 
$H(\Delta,r)$ directly to determine if it is a linear function of $\Delta$. 

For numerical simulations, it is necessary to develop a
model which is simple enough computationally to permit a large inertial range.
Consequently, a one-dimensional scalar field is preferable. However, a
one-dimensional incompressible velocity field would be quite
dull, so we are forced to choose some other form of mixing that
retains certain traits of an incompressible flow.  The attribute preserved
in this model is that the mixing simply re-arranges the scalar field, without
altering any scalar values.  Any advection of this type will necessarily
be non-local.  The particular method of re-arrangement is based on a picture
of a turbulent flow consisting of many swirling eddies, and is constructed
to induce scaling in the passive scalar structure functions. Because 
the model is one-dimensional, it is best thought of as being analogous
physically to a turbulent pipe flow.  This model is loosely based on the 
'linear eddy model' of Kerstein. \cite{ker91} 

The model is designed to produce structure functions which obey statistical 
equations with the same form as equation \ref{StatEq}.  The dissipation functions 
$J_{2n}$ are identical, so that the closure problem is reproduced, but the 
eddy-diffusity operator $\cal L$ is altered (although it remains a scaling operator). 

The motivation for this particular model is developed as follows:
Imagine that the one-dimensional scalar field is embedded in a plane. 
The advection consists of an eddy in that plane, centered on the
scalar field, which rotates one-half turn.  This maps the scalar field
$T(x)$ onto itself, according to the rule
\begin{equation}
	T(x) \to T(L-x)
\end{equation}
where $L$ is the eddy size (centered on $x = {1 \over 2}L$).  The eddy acts 
only on the region $0 < x< L$; elsewhere nothing happens.  Applying one of these
eddies in each time step $\tau$, with randomly chosen size and position, along with
diffusion, gives a rule for advancing the state of the passive scalar by
one time step, 
\begin{equation}
	T(x,\tau)=T(x,0)+V[T(x,0)]+D\tau \partial_x^2 T(x,0)
\end{equation}
in which the advection operator $V[T]$ is
\begin{equation}
	V[T(x)] \equiv  \left\{ 
	\begin{array}{ll}
	 T(2x_o +L-x)-T(x) & \mbox{if $x_o \le x \le x_o + L$,} \\ 0 & \mbox{otherwise.}
	\end{array}
	\right.
\end{equation}
The model consists of applying this rule many times, with the size L and position 
$x_o$ chosen randomly at each step from appropriate probability distribution 
functions.  All possible $x_o$ in the system have equal probability, but 
the eddy sizes $L$ are chosen according to a scaling probability with scaling index $y$:
\begin {equation}
	P(L) dL = CL^{-y}dL.
\end{equation}
This generates scaling behavior within an inertial range determined
by the smallest and largest possible values of $L$: $L_o <L<L_v$.  In the 
analogy to turbulent pipe flow, $L_v$ would play the role of the pipe diameter
and $L_o$ the role of a viscous cut-off scale (the size of the smallest motions).
The passive scalar structure functions would be expected to exhibit inertial
range scaling between $L_o$ and $L_v$, so long as the dissipation scale is not
larger the $L_o$.  In practice, the constant $D\tau$ is chosen so that the dissipation
scale is roughly equal to $L_o$ (corresponding to a Prandtl number near unity), so that 
the ratio of length scales $L_v \over L_o$ plays the role of the Peclet number
in the model.

To maintain a state of statistical equilibrium, some forcing is required.  In this
model, the forcing is done by imposing an overall gradient $g$ on the scalar, as in \cite{shr95}. A new variable $\theta$ is then defined by the deviation from the gradient, 
\begin{equation}
	\theta (x) \equiv T(x) - gx
\end{equation}
and structure functions are defined in terms of $\theta$:
\begin{equation}
	S_n(r) \equiv \langle (\theta (x+r) - \theta (x))^n \rangle .
\end{equation}

\subsection{Structure Function Equations}
Equations for the structure functions $S_n(r)$ can be constructed in the same 
way as in Section 1.2.  The scalar difference 
$\Delta (x,y) \equiv \theta (x) - \theta (y)$ obeys the equation
\begin{equation}
	\Delta (x,y;\tau) = D\tau (\partial_x^2 + \partial_y^2 ) \Delta (x,y;0)+
	\Psi [\Delta(x,y;0)]
\end{equation}
when time is advanced by one unit $\tau$.  The action of the convective term $\Psi$
on $\Delta$ depends on whether x or y (or both) lie within the eddy:
\begin{equation}
        \Psi [\Delta(x,y)] \equiv  \left\{
        \begin{array}{ll}
         	\Delta (x,y) & x,y \notin [x_o,x_o +L] \\ 
		\Delta (2x_o +L-x,y) + g(L+2x_o -2x) & 
			x \in [x_o,x_o +L], \\
			& y \notin [x_o,x_o +L] \\
		\Delta (x,2x_o +L-y) - g(L+2x_o -2y) &
                        x \notin [x_o,x_o +L], \\ 
			& y \in [x_o,x_o +L] \\
		\Delta (2x_o +L-x,2x_o +L-y) - 2g(x-y) & x,y \in [x_o,x_o +L].	
        \end{array}
        \right.
\end{equation}
Raising this equation to the $n$th power and taking the ensemble average gives
\begin{equation}
	S_n (x-y;\tau ) = \langle \Psi ^n [\Delta(x,y;0)] \rangle +
	nD\tau \langle \Psi ^{n-1} [\Delta(x,y;0)] (\partial_x^2 +  \partial_y^2 ) 
	\Delta (x,y;0) \rangle +...
\end{equation}
keeping only the lowest order term in $D\tau$.  In the diffusive term, $\Psi [\Delta]$
can be replaced with $\Delta$, because only a small portion of the scalar field lies
within the eddy at that time step (the eddy is another higher order correction).
Then the equation becomes
\begin{equation}
	S_n(x-y;\tau ) = \langle \Psi^n [\Delta(x,y;0)] \rangle + J_n(x,y;0) 
\end{equation}
where
\begin{equation}
	J_n(x,y) \equiv nD\tau \langle \Delta(x,y)^{n-1} 
	(\partial_x^2 + \partial_y^2 ) \Delta(x,y) \rangle
\end{equation}
as in equation \ref{JDef} and  \cite{kra94}. The quantity $\langle \Psi ^n [\Delta(x,y;0)] \rangle$ 
can be computed from the definition of $\Psi$ by taking the ensemble average
over the eddy probabilities for $x_o$ and $L$. Defining the difference variable 
$r \equiv x-y$, the result divides into three regions: a diffusive interval 
$(r \le L_o)$, the inertial range $(L_o \le r \le L_v)$, and a large scale region 
$(r \ge L_v)$. The structure functions obey the equation 
\begin{equation}
	\Lambda(S_n(r,\tau)-S_n(r,0)) +{\cal L}[S_n(r,0)] = \Lambda J_n(r,0) +F_n(r,0)
\end{equation}
where $\Lambda$ is the system size. In statistical equilibrium $S_n(r,\tau)=S_n(r,0)$,
and this equation is identical in form to equation \ref{StatEq}.  The eddy-diffusivity 
operator $\cal L$ and the source $F_n(r)$ differ according to the value of $r$.  
For $r \le L_o$, they are 
\begin{equation}
        {\cal L}[S(r)] \equiv  -{\int_{L_o}^{L_v} P(L) dL} \{
		{\int_{L-r}^{L+r} dz S(z)} -(r+L)S(r)-(r-L)S(-r) \}
\end{equation}
\begin{eqnarray}
	F_n(r) \equiv \sum_{m=1}^n {g^m n! \over m!(n-m)!} {\int_{L_o}^{L_v}
        	P(L) dL} \{ {\int_{L-r}^{L+r} dz S_{n-m}(z)(z-r)^m} \nonumber \\ 
       		+ (L-r)S_{n-m}(-r)(-2r)^m \}.
\end{eqnarray}
In the inertial range, $L_o \le r \le L_v$, they are
\begin{eqnarray}
	{\cal L}[S(r)] \equiv -{\int_{L_o}^{r} P(L) dL \{    
        	{\int_{r-L}^{r+L} dz S(z)} -2LS(r) \} } \nonumber \\ - {\int_{r}^{L_v}
        	P(L) dL \{ {\int_{L-r}^{L+r} dz S(z)} -(r+L)S(r) -(r-L)S(-r) \} }
\end{eqnarray}
\begin{eqnarray}
	F_n(r) \equiv \sum_{m=1}^n {g^m n! \over m!(n-m)!} \{{\int_{L_o}^{r} P(L) dL}
        	{\int_{r-L}^{r+L} dz S_{n-m}(z)(z-r)^m} \nonumber \\
        	+ {\int_{r}^{L_v} P(L) dL} \{ {\int_{L-r}^{L+r} dz S_{n-m}(z)(z-r)^m}
        	+ (L-r)S_{n-m}(-r)(-2r)^m \} \}.	
\end{eqnarray}
In the large scale region, $r \ge L_v$, they are
\begin{equation}
	{\cal L}[S(r)] \equiv -{\int_{L_o}^{L_v} P(L) dL \{
        	{\int_{r-L}^{r+L} dz S(z)} -2LS(r) \} }
\end{equation}
\begin{equation}
	F_n(r) \equiv \sum_{m=1}^n {g^m n! \over m!(n-m)!}
	        {\int_{L_o}^{L_v} P(L) dL} \{ {\int_{r-L}^{r+L} dz
       		S_{n-m}(z)(z-r)^m} \}.
\end{equation}

\subsection{Solution for $S_2(r)$}

The first order structure function $S_1$ vanishes, so the lowest non-trivial structure
function is $S_2$.  It can be determined approximately in both the inertial and
dissipative ranges.  As a first step, it is necessary to re-write $J_2$ by commuting 
derivatives and using spatial homogeneity, as in the Kraichnan model:
\begin{equation}
	J_2 = 2D\tau \langle \Delta (\partial_x^2+\partial_y^2)\Delta \rangle 
	= 2D\tau \partial_r^2 S_2 - 4D\tau \langle (\partial_x\theta)^2 \rangle
\end{equation}
where $D\tau \langle (\partial_x\theta)^2 \rangle$ is the mean-square dissipation of 
the scalar (a constant).  This constant can be evaluated by looking at the large scale 
region (where $S_2$ must approach a constant)
and balancing it with the forcing term $F_2 = {2\over3} g^2 \langle L^3 \rangle$, so that 
\begin{equation}
	\langle (\partial_x \theta)^2 \rangle 
		= {1\over6} {\langle L^3 \rangle\over D\tau\Lambda} g^2.
\end{equation}
This result serves as a check on the accuracy of numerical simulations. 

The solution far into the dissipative region $(r\ll L_o)$ can then be evaluated
by neglecting both $\cal L$ and $F$ (setting $J_2 =0$):
\begin{equation} \label{S2D}
	S_2(r) \cong \langle (\partial_x\theta)^2 \rangle r^2 
		= {1\over6} { \langle L^3 \rangle \over{D\tau\Lambda}}(gr)^2.
\end{equation}

In the inertial range, the approximate solution is found by balancing the convective 
term $\cal L$ against the dissipation term 
$J_2 \cong -4D\tau \langle (\partial_x\theta)^2 \rangle$:
\begin{equation}
	{\cal L} [S_2(r)] \cong  -{2\over3} \langle L^3 \rangle g^2. 
\end{equation} 
By assuming a scaling solution, $S_2(r)=A_2r^{\rho_2}$, the convective term becomes
\begin{equation}
	{\cal L} [A_2r^{\rho_2}] \cong -{A_2 L_o^y \over {y-1}} I(\rho_2,y)r^{2+\rho_2-y}
\end{equation}
in the limits ${L_o\over r}\to 0$ and ${r \over L_v}\to 0$. The definite integral 
$I(\rho,y)$ is defined as
\begin{eqnarray}
	 I(\rho,y) \equiv {1\over 1+\rho} \{ {\int_0^1
        {dz\over z^y}((1+z)^{1+\rho}-(1-z)^{1+\rho}-2(1+\rho)z)} \nonumber \\ 
	+ {\int_0^1 {dz\over z^{3+\rho-y}}((1+z)^{1+\rho}-(1-z)^{1+\rho}
	- 2(1+\rho)z^{1+\rho})} \}
\end{eqnarray}
and can be evaluated numerically. The first integral in $I$ diverges at the lower 
limit for $y>3$, setting an upper limit on the scaling range. (For
$y>3$ the lower limit must be replaced by ${L_o\over r}$, and the ultraviolet region 
determines the solution.) 
Since ${\cal L}[S(r)]$ balances a constant, the scaling is fixed at $\rho_2=y-2$ 
(which only makes sense for $y>2$) and the solution for $L_o \ll r \ll L_v$ is 
\begin{equation} \label{S2R}
	S_2(r)={2\over3} {(gL_v)^2 \over (2-\rho_2)I(\rho_2,y)} 
		\left( {r\over L_v} \right)^{\rho_2} 
\end{equation}
where $\langle L^3 \rangle$ has been evaluated explicitly.

These solutions for $S_2$ compare very well with the numerical results in both the 
inertial and dissipative ranges (see Figure \ref{fig5}).

\subsection{Higher Order Structure Functions}
\subsubsection*{Even-Order Structure Functions}

In the inertial range, the scaling of higher even-order structure functions is 
determined by balancing ${\cal L} [S_{2n}]$ against $J_{2n}$, as in \cite{kra94}:
\begin{equation} \label{LeqJ}
	{\cal L} [S_{2n}(r)] \cong J_{2n}(r).
\end{equation}
The closure assumption for $J_{2n}$ is that 
\begin{equation} \label{Jn}
	J_{2n}(r) = nC_n J_2{S_{2n} \over S_2}.
\end{equation}
This generalized form with $C_n \ne 1$ was suggested in \cite{chi96}, while $C_n =1$
gives the original Kraichnan ansatz of \cite{kra94}.
Assuming scaling functions for $S_{2n}$, the scaling exponents $\rho_{2n}$
are determined by the coefficients in equation \ref{LeqJ}:
\begin{equation}
	I(\rho_{2n},y)=nC_nI(\rho_2,y)
\end{equation}
This equation for the exponents is analogous to equation \ref{ModExp} in the Kraichnan model. 
This result can be used (in principle) to evaluate the coefficients of proportionality 
$C_n$ given numerically measured $\rho_{2n}$.  Unfortunately, the sensitivity of
$I$ to $\rho$ limits the accuracy of this method.  However, the numerical results 
for the exponents and $C_n$ (computed from equation \ref{Jn}) are consistent with this result, 
indicating that the scaling is indeed set by
balancing the convective term $\cal L$ against the dissipation term $J_{2n}$.

\subsubsection*{Odd-Order Structure Functions}

Odd order structure functions exhibit scaling only in the dissipative range.  
Because the eddy-diffusivity operator $\cal L$ differs depending on whether it 
operates on an odd or even function of $r$, the inertial range scaling
solution exists only for even order structure functions.  Odd order functions are 
positive scaling functions in the dissipation range (small $r$), 
but pass through zero in the inertial range and then approach zero from below at large $r$. 
This behavior of the odd structure functions is peculiar to this particular model 
and differs from physical passive scalars.

\subsection{Dissipation Range Scaling}

At small enough length scales the dissipative terms determine the scaling.  In this
(dissipative) range the $J_n$ term dominates the solution.  It is convenient to re-write it
as:
\begin{equation}
	J_n (r)=2D\tau\partial_r^2 S_n(r)-
	n(n-1)D\tau \langle \Delta^{n-2}[(\partial_x\theta)^2+(\partial_y\theta)^2] \rangle .
\end{equation}
Balancing the two parts of $J_n$ against each other will lead to a solution if the conditional
probability
\begin{equation}
	G(\Delta,x-y) \equiv \langle [(\partial_x\theta)^2+(\partial_y\theta)^2]|\Delta \rangle
\end{equation}
can be calculated. In a purely dissipative system, $G$ can be determined from a Taylor
expansion of $\Delta$ at small separations: $\Delta \cong r(\partial_x\theta)$, which implies 
$G = {2\Delta^2 \over r^2}$.  However, the presence of convection 
competes with the smoothing effects of diffusion and generates higher terms in the Taylor 
expansion of $\Delta$.  Given this fact, a reasonable
closure approximation (which is supported numerically) is
\begin{equation}
	G(\Delta,r) = a + {b\Delta \over r} + {c\Delta^2 \over r^2}
\end{equation}
where $a$, $b$, and $c$ are constants.  The assumed $r$ dependence is necessary to generate a
solution with regular scaling.  

For a regular scaling solution, $S_n(r)=A_n r^n$, the unknown constants can be expressed in
terms of $A_2$ (calculated above) and $A_3$ (unknown):
\begin{equation}\label{Dis}
	A_n = A_{n-2} A_2 + {A_3 \over A_2} A_{n-1}.
\end{equation}
Hence all higher order structure functions are expressible in terms of $S_2$ and $S_3$. 
Unfortunately, no analytic solution for $S_3$ has been found.  However, in the absence of
any spatial assymmetry (which is generated by the gradient forcing used in this model)
the odd order structure functions would vanish and the solution would be
\begin{equation}
	S_{2n}(r)=S_2^n (r)
\end{equation}
which has regular scaling but non-gaussian statistics.  

\section{Numerical Results}
\subsection{Simulation of the Model}

A computational model was created by discretizing the dynamical equation for $T(x)$. 
The advective term is handled straightforwardly, since it is just a re-arrangement
of the values of $T(x)$.  Diffusion was done subsequent to the advection process 
in each time step, using a second-order implicit finite differences scheme. 
An overall gradient was applied to generate forcing, and periodic boundary
conditions for the fluctuating field $\theta (x)$ were used.  The system 
was evolved to a state of statistical equilibrium before any averaging computations 
were done.  Equilibrium was indicated by the establishment of stable (analytically 
known) values of $\langle (\partial_x \theta)^2 \rangle$ and $S_2(r)$.  Structure 
functions were computed by taking space-time averages over the entire simulation, 
implicitly assuming an ergodic system.  
\begin{figure}[htb]                                                      
\centerline{\epsfbox{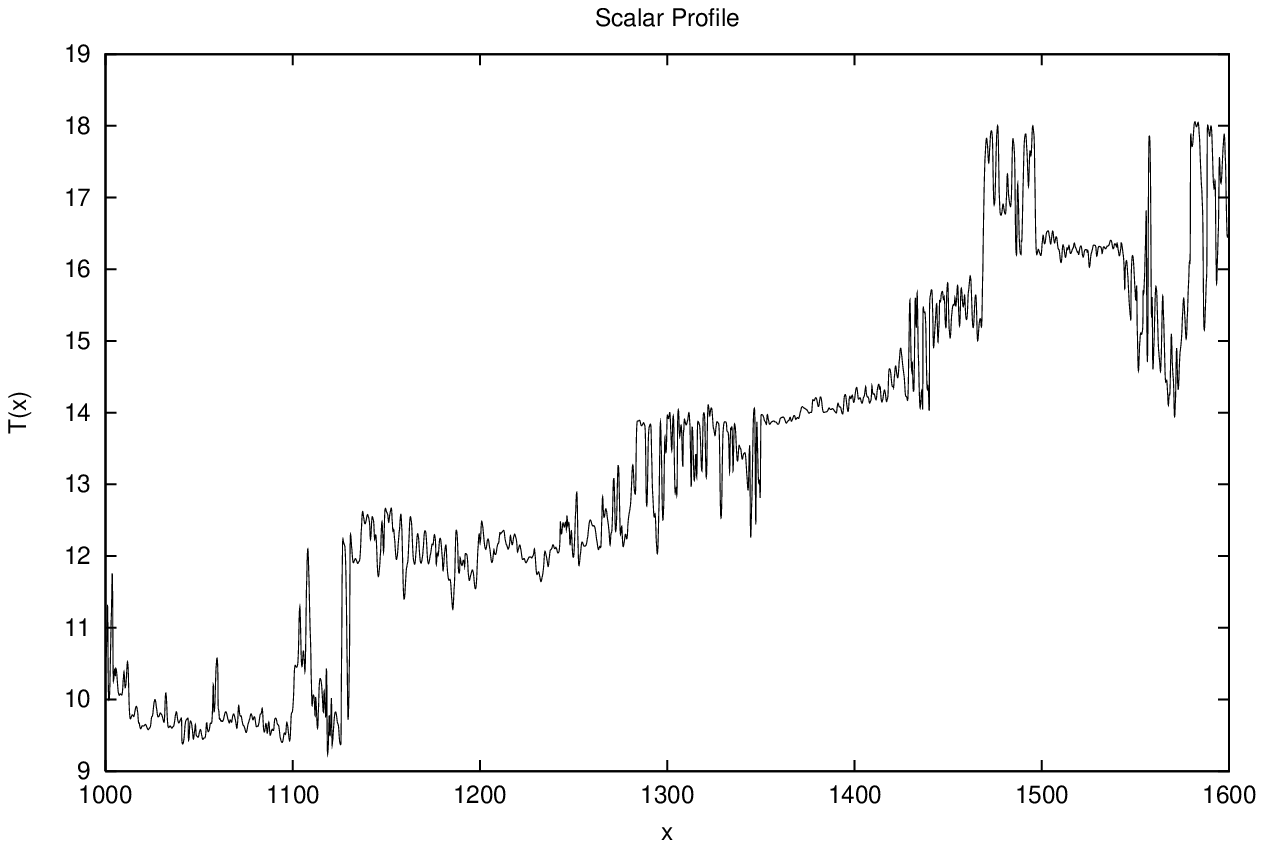}}                                       
\small{
A `snapshot' of the scalar profile, revealing the overall gradient
$g=0.01$ along with significant fluctuations.  The profile represents 6000
grid points ($dx=0.1$).}
\caption{Typical Scalar Profile}
\bigskip
\label{fig2}                                                
\end{figure}            

Simulations were conducted in a system with 50000 grid points and a grid spacing of
$dx=0.1$.  The smallest eddy size was $L_o=2$, permitting a significant 
dissipative interval.  The largest eddy size was either $L_v=200$ or  
$L_v=500$, allowing for inertial range scaling over two orders of magnitude.
The imposed gradient was $g=0.01$, and the diffusion constant $D\tau$ was of
order $10^{-5}$ (varying slightly with $y$ to set the dissipation scale approximately
equal to $L_o$).  The eddy scaling exponent was varied between $y=2.1$ and $y=2.8$ 
for $L_v=200$, and between $y=2.1$ and $y=2.4$ for $L_v=500$. The number of time
steps varied between $3\cdot 10^7$ and $10^8$.  A large number
of time steps is needed to suitably average over the eddy size probability 
distribution $P(L)$, especially for larger $L_v$ and $y$.  In addition, several 
simulations were conducted with large diffusion constants ($D\tau = 0.02$) to study 
the dissipative range solution. A typical `snapshot' of the scalar profile
$T(x)$ is shown if Figure \ref{fig2}. The profile reveals structure on a wide range of
length scales, including significant `flat' regions, and appears similar to
experimentally observed profiles.  

\subsection{Numerically Computed Structure Functions}

\begin{figure}[htb]
\centerline{\epsfbox{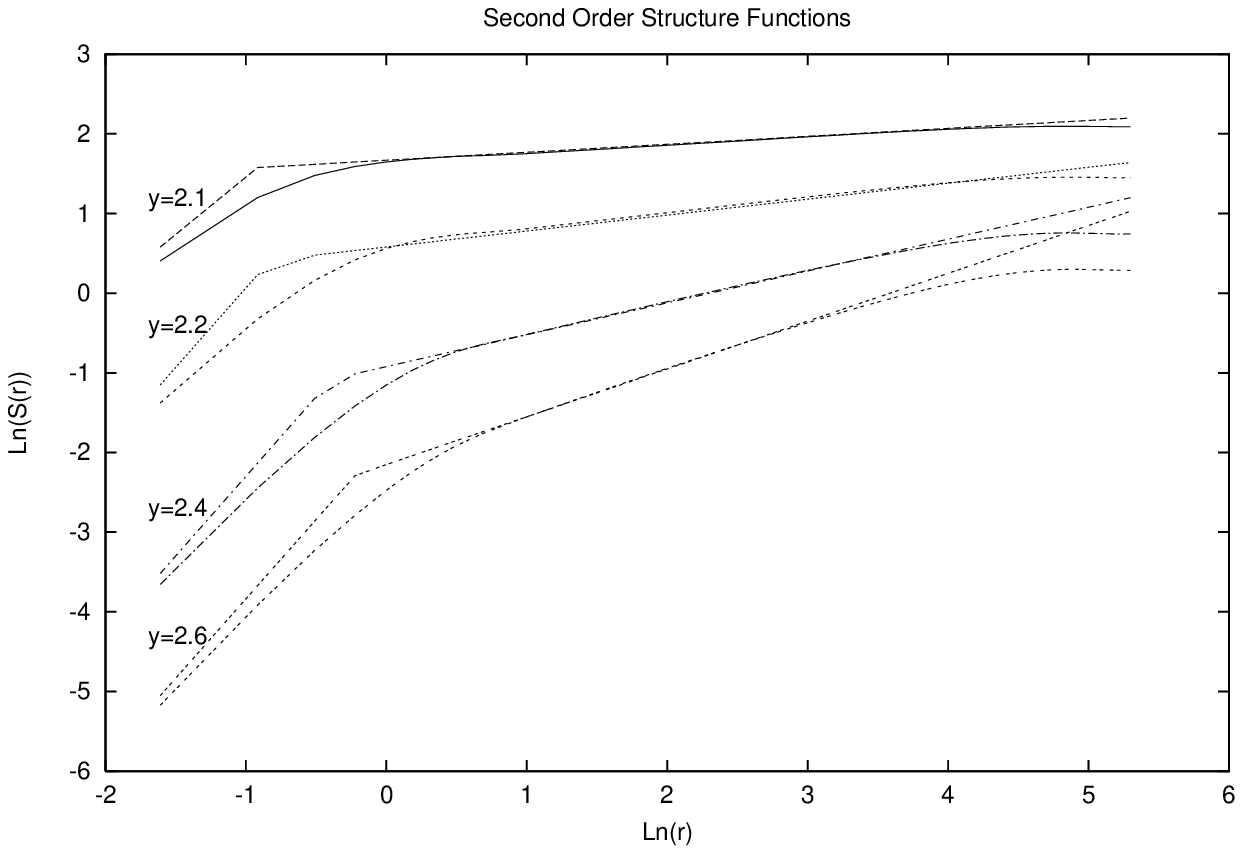}}
\small{
Second order structure functions $S_2(r)$ on a log-log plot, for
$L_v=200$ and four values of the scaling parameter $y$.  Smooth curves are
numerical simulation data;  straight lines represent the analytic scaling solution
in two regimes (inertial and dissipative ranges).}
\caption{Second Order Structure Functions}
\label{fig5}
\bigskip
\end{figure}
Second order structure functions $S_2$ have been computed analytically and can be 
compared directly with simulation results.  Figure \ref{fig5} shows $S_2$ as a function 
of $r$ on a log-log plot (base $e$) for $L_v=200$ and four values of $y$.  The smooth 
curves show the simulation data (50 data points each).  The analytic results are 
shown as two straight lines (equation \ref{S2D} for the dissipative region, and 
equation \ref{S2R} for the inertial range).  There are no free parameters. 
The agreement is good (within 5\% in the heart of the inertial range) except at
the inertial range boundaries, where the analytic calculation has no validity.  
Interestingly, the boundary terms apparently grow in importance as $y$ increases,
and this encroachment reduces the effective size of the inertial range.

\begin{figure}[htb]
\centerline{\epsfbox{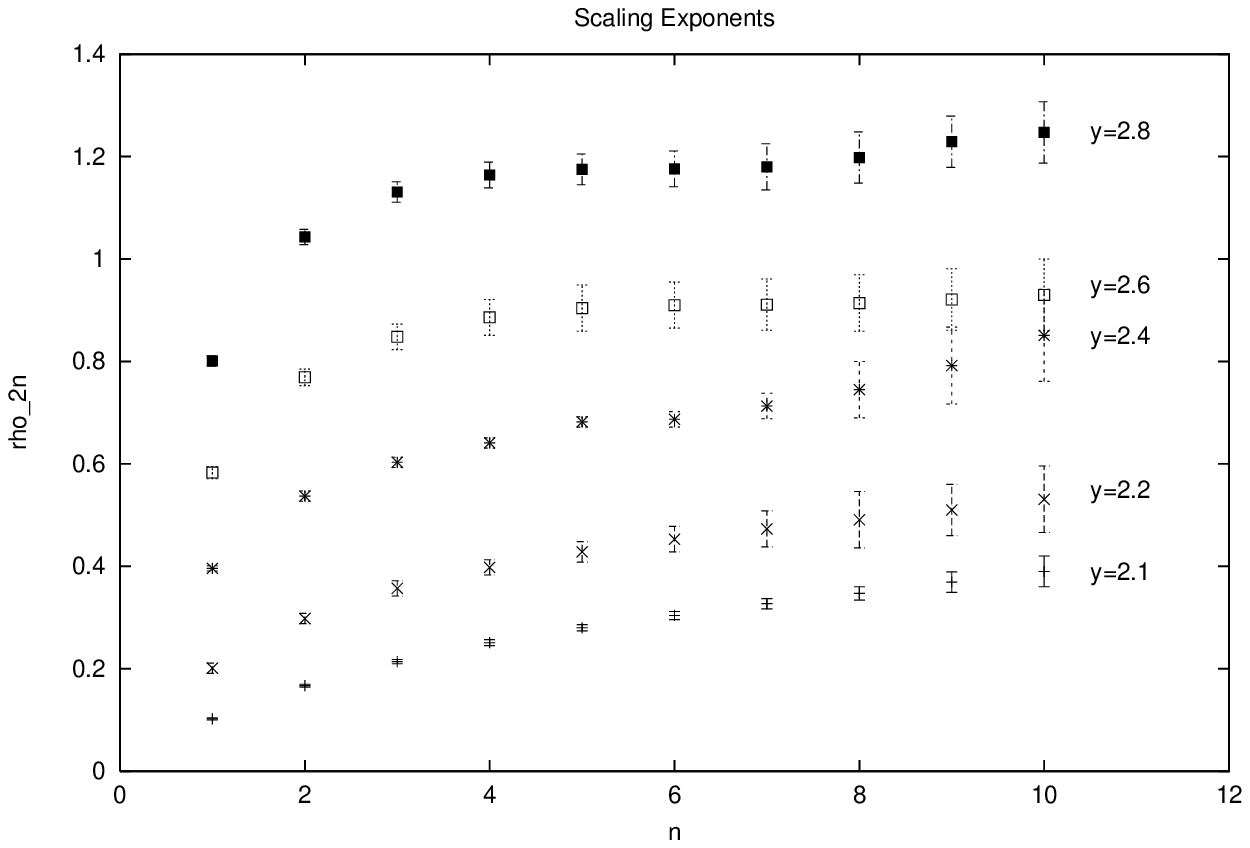}} 
\small{
Inertial range structure function scaling exponents $\rho_{2n}$
as a function of $n$ for five values of $y$. These simulation data agree with
the analytically known result $\rho_2 = y-2$ for $n=1$.  The data suggest that
the exponents may approach a constant value as $n \to \infty$.}
\caption{Inertial Range Scaling Exponents}
\bigskip
\label{fig7}
\end{figure}

Higher even-order structure function also exhibit inertial range scaling.
Figure \ref{fig7} shows the even order scaling indices $\rho_{2n}$ as a function of $n$ 
for five values of $y$.  The error bars represent the range of observed values over
several simulations with different initial conditions.  The scaling indices are 
independent of the upper length scale $L_v$ (within the error). The second 
order indices $\rho_2$ lie within 3\% of the theoretical values $y-2$.  The deviation
from regular scaling $(\rho_{2n}=n\rho_2)$ is quite pronounced.  
For larger $y$, the scaling exponents appear to approach a constant value 
(dependent on $y$) as $n$ increases.
\clearpage

\subsection{Probability Distribution Functions}
\begin{figure}[htb]
\centerline{\epsfbox{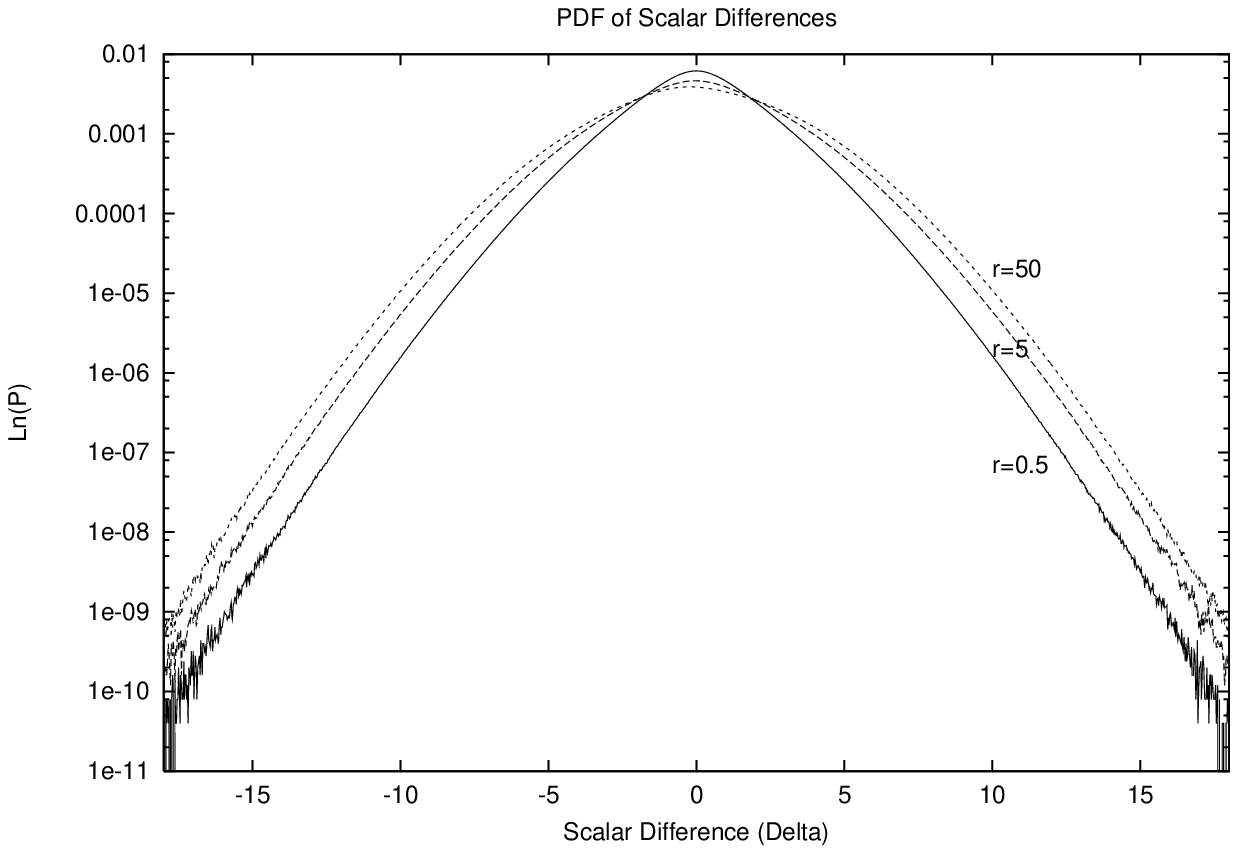}} 
\small{
Probability distribution function $P(\Delta)$ of scalar
differences $\Delta$ for three separations $r$, from the $y=2.1$, 
$gL_v=2$ simulation.  The tails ($|\Delta| > gL_v$) of the distribution 
appear to be exponential, with a slope which is independent of $r$.}
\caption{Scalar Difference PDF, $y=2.1$}
\bigskip
\label{fig8}
\end{figure}

\begin{figure}[htb]                            
\centerline{\epsfbox{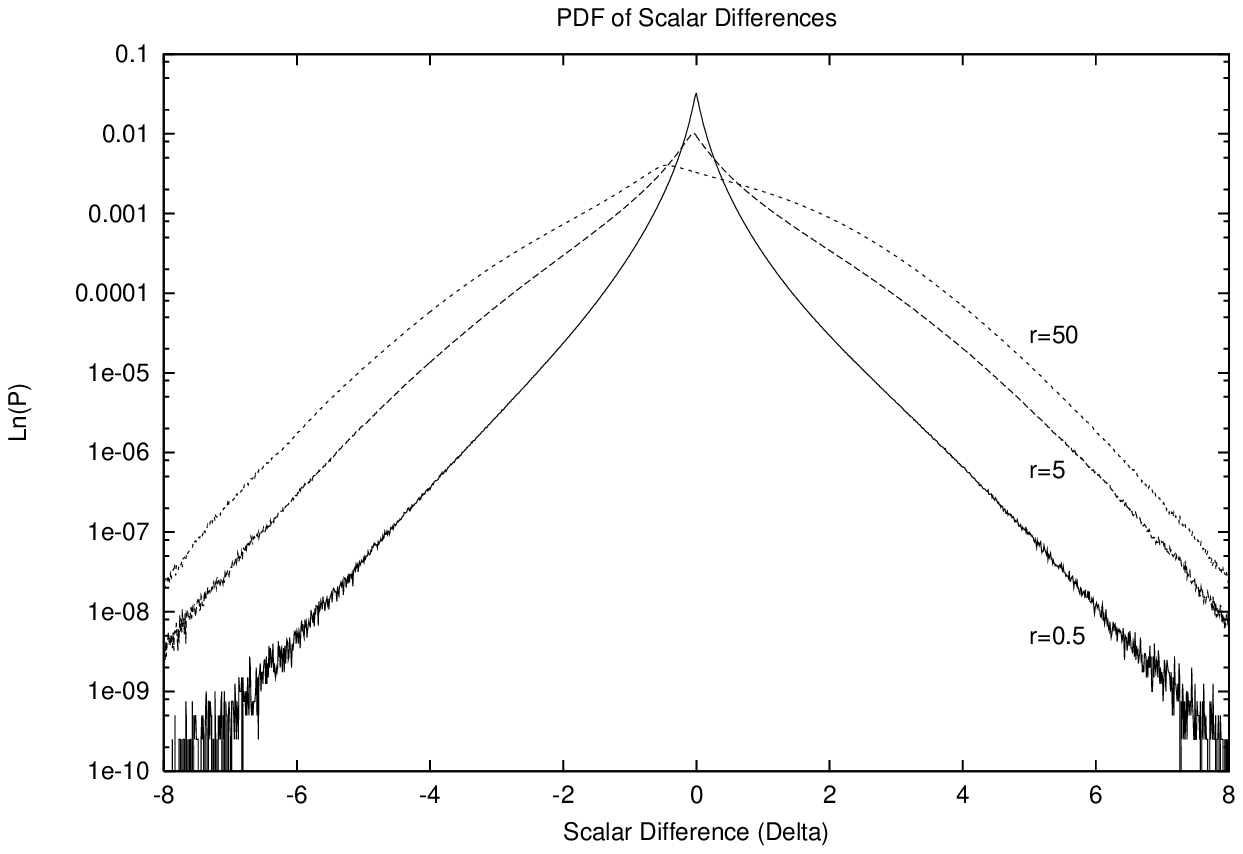}}                             
\small{
Probability distribution function $P(\Delta)$ of scalar
differences $\Delta$ for three separations $r$, from the $y=2.4$, 
$gL_v=2$ simulation.  The tails ($|\Delta| > gL_v$) of the distribution            
appear to be exponential, with a slope which is independent of $r$.}
\caption{Scalar Difference PDF, $y=2.4$}
\bigskip
\label{fig9}
\end{figure}

Figure \ref{fig8} shows the probability distribution function (PDF) for $\Delta$ 
on a log-linear scale for several values of $r$ from the $y=2.1$, $L_v=200$ simulation.
The core of the PDF, defined roughly by $|\Delta| \leq gL_v$ ($gL_v=2$), is rounded.  
The slight asymmetry between positive and
negative $\Delta$ is due to the imposed gradient.  Figure \ref{fig9} shows the
PDFs for the same values of $r$ from the $y=2.4$, $L_v=200$ simulation.  
The core of this PDF is much more sharply peaked.

All the numerically generated PDFs exhibit exponential tails for $|\Delta| \gg gL_v$:
\begin{equation}
	P(\Delta ,r) \cong A(r) e^{-c|\Delta |}
\end{equation}
where $c$ is independent of the separation $r$.  Exponential tails have also 
been derived in another model in a particular limit. \cite{che97}  This form of the PDF
suggests that, for large $n$, the structure function scaling exponents $\rho_{2n}$ 
approach a constant independent of $n$.  In this case the structure functions obey
\begin{equation}
	\lim_{n \to \infty} {S_{2n} \over S_{2n-2}} = {2n(2n-1) \over c^2}
\end{equation}
in the limit of large $n$.  The PDF's and the structure functions can both be 
used to independently estimate $c$, and the results are shown in Figure 3.9
(as a function of $\rho_2$).  The two sets of data 
represent the two different values of $L_v$, $L_v=200$ and $L_v=500$. 

The exponential tails can be understood in terms of a random walk of fluid 
elements.  To generate a particular temperature difference $\Delta$, fluid elements 
initially separated by a distance of order $\Delta \over g$ must be brought close 
together.  Since the largest correlated motion in the system is of
size $L_v$, scalar differences larger in magnitude than $gL_v$ can only be 
generated by the uncorrelated action of several eddies.  This behavior is essentially 
a random walk, and the multiplication of probabilities leads to the exponential PDF.

\begin{figure}[htb]
\centerline{\epsfbox{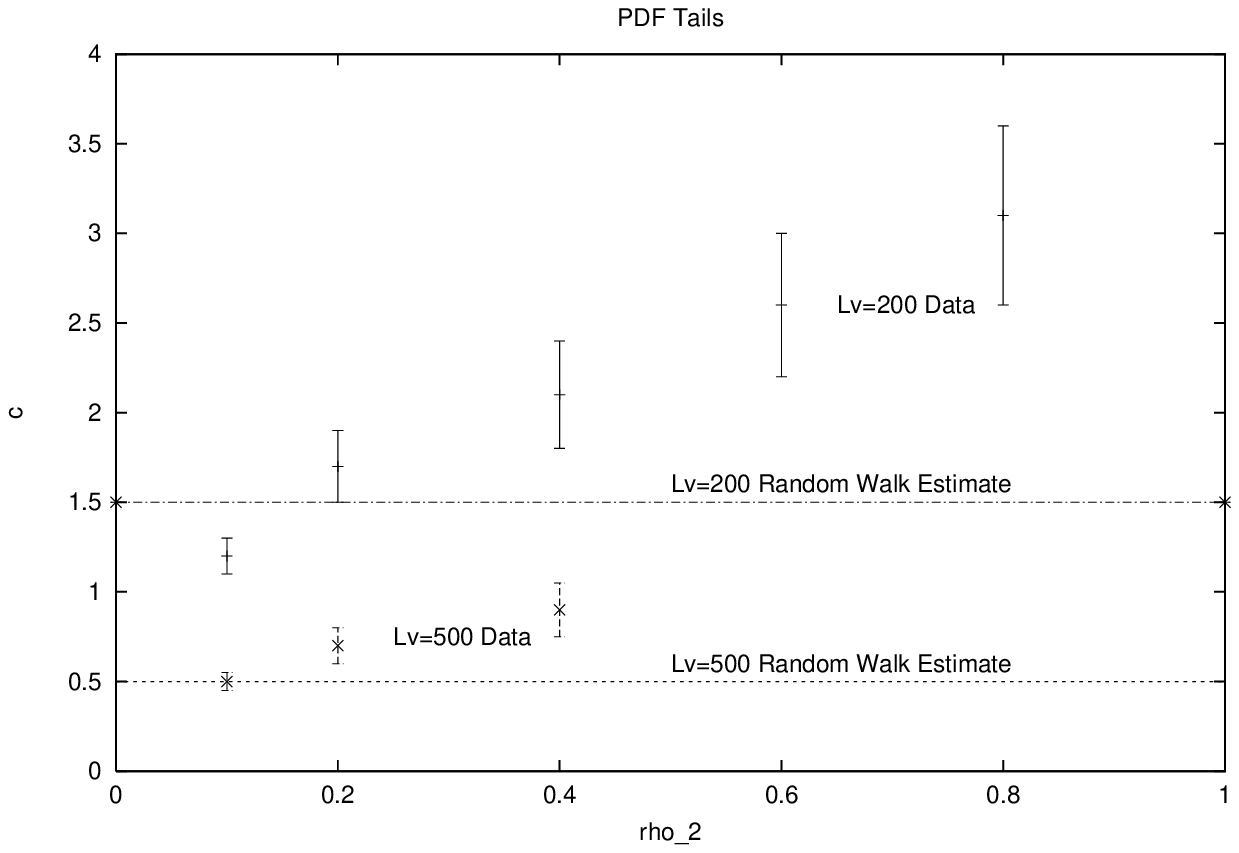}} 
\small{
Slopes $c$ of the tails of the PDF $P(\Delta)$ measured from
simulation data for $L_m=200$ and $L_m=500$.  The straight lines represent the
Lagrangian random-walk estimate of $c$.  The dependence on $\rho_2$ arises because
large eddies become less frequent as $\rho_2$ increases.}
\caption{Slopes of PDF Tails}
\bigskip
\label{fig10}
\end{figure}

To be more quantitative, consider the probability for the motion of a fluid 
element along a Lagrangian trajectory from position zero
to position $x$.  For simplicity assume that all eddies are the same size $L$.  
To move a distance $x$ with $(m-1)L \leq x \leq mL$ (for integer $m$) will 
typically require the point to be moved by $m$ eddies, each carrying it
a distance of order $L$.  However, the probability that the point will lie within a 
particular eddy is $L \over \Lambda$, and these probabilities multiply:
\begin{equation}
	P(x \sim mL) \sim \left( {L \over \Lambda} \right) ^m. 
\end{equation}
The motion must occur quickly, in (order of magnitude) $m$ steps, or else diffusion 
will cause the fluid element to equilibrate with its new environment.
By assuming the two points involved in constructing $\Delta$ move independently, 
this probability can be converted to a PDF for $\Delta$ by using 
$x \sim {\Delta \over g}$ regardless of the separation $r$.  The result is
\begin{equation}
	 P(\Delta) \sim e^{-c \Delta}
\end{equation}
\begin{equation}
	 c \equiv {1 \over gL} ln \left( {\Lambda \over L} \right) .
\end{equation}
Assuming that each eddy is of size $L_v$ suggests estimates of $c=1.5$ for 
$L_v=200$ and $c=0.5$ for $L_v=500$.  These values are shown as straight lines in 
Figure \ref{fig10}, and are in qualitative agreement with the data. 
Of course, not all eddies are of size $L_v$;  in fact the number of
large eddies decreases as $\rho_2$ increases.  This results in the trend of 
increasing $c$ shown in Figure \ref{fig10}.

Once two particles have come close together, they become subject to correlated 
motions and diffusion.  This is presumably the source of the $r$ dependence in the 
pre-factor $A(r)$ of the exponential tail. Correlated motions have a weak influence on 
the PDF tail, and cannot alter its exponential character. Also, there is a slight 
asymmetry between $\Delta$ and $-\Delta$ in the prefactor;  this is attributed to the
fact that producing a negative $\Delta$ requires the two points to pass by each other
(in one dimension), and during this time their motion is correlated. 

\subsection{Kraichnan's Closure Ansatz}

To test the closure ansatz in \cite{kra94} and evaluate the constants of 
proportionality $C_n$, the dissipation functions $J_{2n}$ were computed directly 
from the simulation data. Rather than use the Laplacian, $J_{2n}$ was re-written 
by commuting derivatives as
\begin{equation}
	J_{2n}(r)=2D\tau\partial_r^2 S_{2n}(r)-2n(2n-1)D\tau \langle \Delta^{2n-2}
	[(\partial_x\theta)^2+(\partial_y\theta)^2] \rangle .
\end{equation}
In the limit $D\tau \to 0$, only the second term on the right side remains.  In
actual simulations, the first term makes a finite contribution which makes it 
more difficult to determine the inertial range scaling.  So in practice only the
second term (approximated using finite differences) was used as a surrogate for 
$J_{2n}$ in the inertial range. 
\begin{figure}[htb]
\centerline{\epsfbox{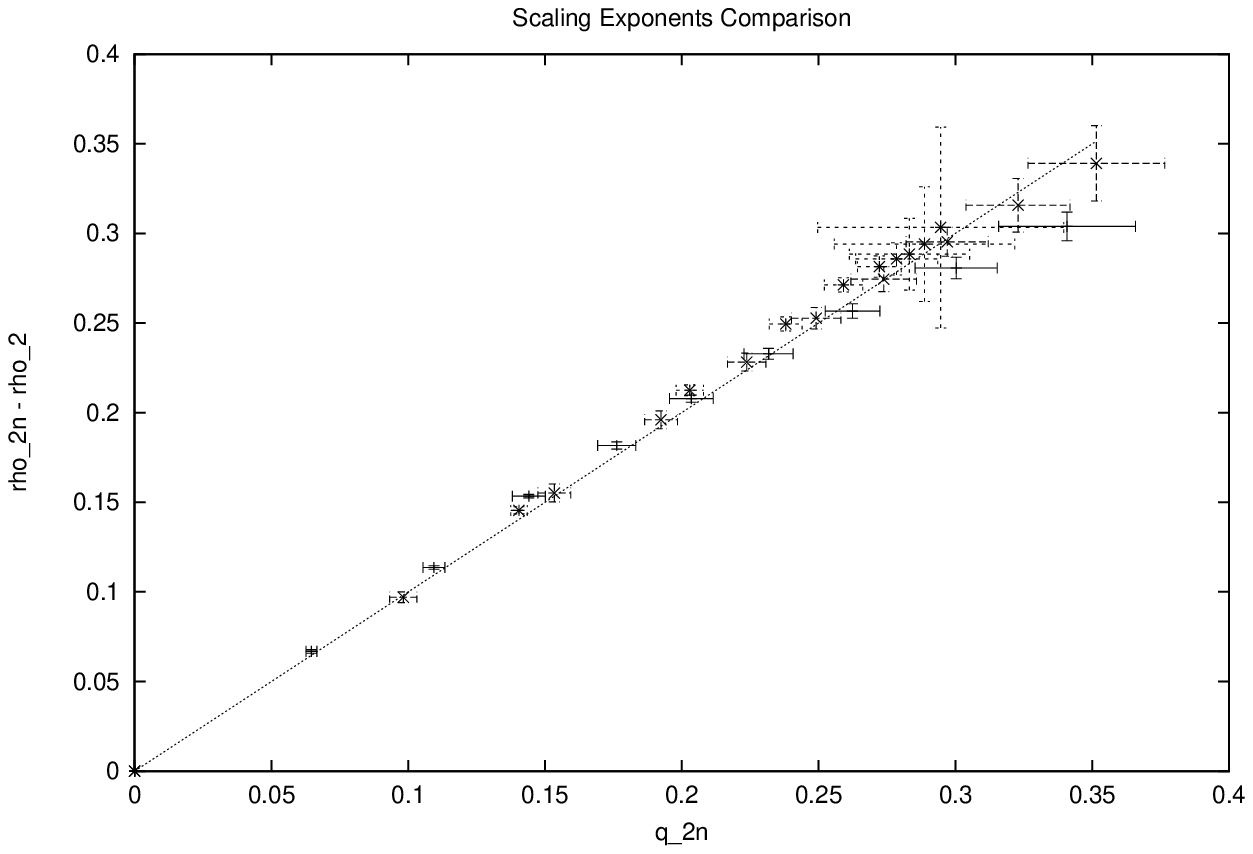}}
\small{
Inertial range scaling exponents $q_{2n}$ for the dissipation
functions $J_{2n}$ compared to the structure function scaling exponents
$\rho_{2n}$.  Kraichnan's closure ansatz requires $q_{2n}=\rho_{2n}-\rho_2$,
and the simulation data is consistent with this result.}
\caption{Scaling Exponent Comparison}
\bigskip
\label{fig12}
\end{figure}

In the inertial range, the $J_{2n}$ are scaling functions of $r$ with scaling indices 
$q_{2n}$ which, by the closure ansatz, ought to satisfy 
\begin{equation}
	q_{2n}=\rho_{2n}-\rho_2.
\end{equation}
Figure \ref{fig12} shows that this part of the closure ansatz holds, by plotting 
$q_{2n}$ vs. $\rho_{2n}-\rho_2$ for the $L_v=500$ simulations.
\begin{figure}[htbp]
\centerline{\epsfbox{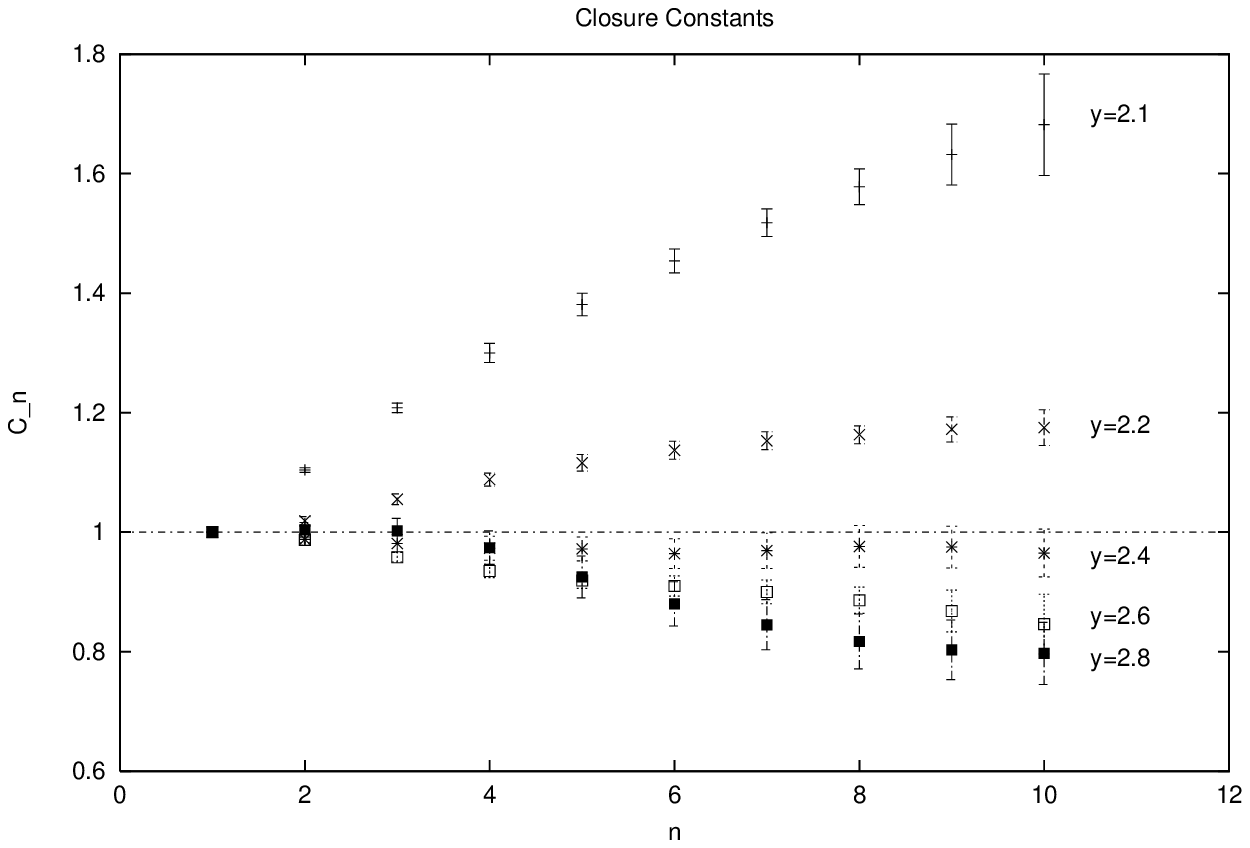}}
\small{
Closure constants of proportionality $C_n$, defined by
$J_{2n} = nC_n J_2{S_{2n} \over S_2}$, from simulation data for several $y$.
Kraichnan's closure ansatz requires $C_n=1$, which is inconsistent with the data.
For larger $y$, $C_n$ decreases with $n$, suggesting scaling exponents might
approach a constant value at large $n$.  For smaller $y$, $C_n$ increases with
$n$, suggesting near-regular scaling for moderate values of $n$.  Both trends
are consistent with the scaling exponents shown in Figure \ref{fig7}.}
\caption{Closure Constants of Proportionality}
\label{fig13}
\end{figure}

The constants $C_n$ can be computed using $C_n={J_{2n}S_2 \over nJ_2S_{2n}}$ and 
averaging over the inertial range ($C_n$ is approximately constant over the region 
averaged). The resulting values to not appear to depend on any parameters of 
the system other than $y$ (a weak dependence on
$L_v$ is suspected but not detectable).  Figure \ref{fig13} shows computed values
for $C_n$ averaged over several simulations.  The deviation from $C_n=1$ is small 
but significant, contradicting the ansatz 
of \cite{kra94}.  The growing values of $C_n$ at small $\rho_2$ suggests near-regular
scaling in this limit, at least for moderate values of $n$.  This is consistent
with the scaling exponents in Figure \ref{fig7}.   The intermediate case $\rho_2=0.4$
lies very close to the Kraichnan prediction of $C_n=1$.  For larger $\rho_2$, $C_n$
decreases with $n$, consistent with the scaling exponents approaching a constant 
value (as in Figure \ref{fig7}).
\begin{figure}[htbp]
\centerline{\epsfbox{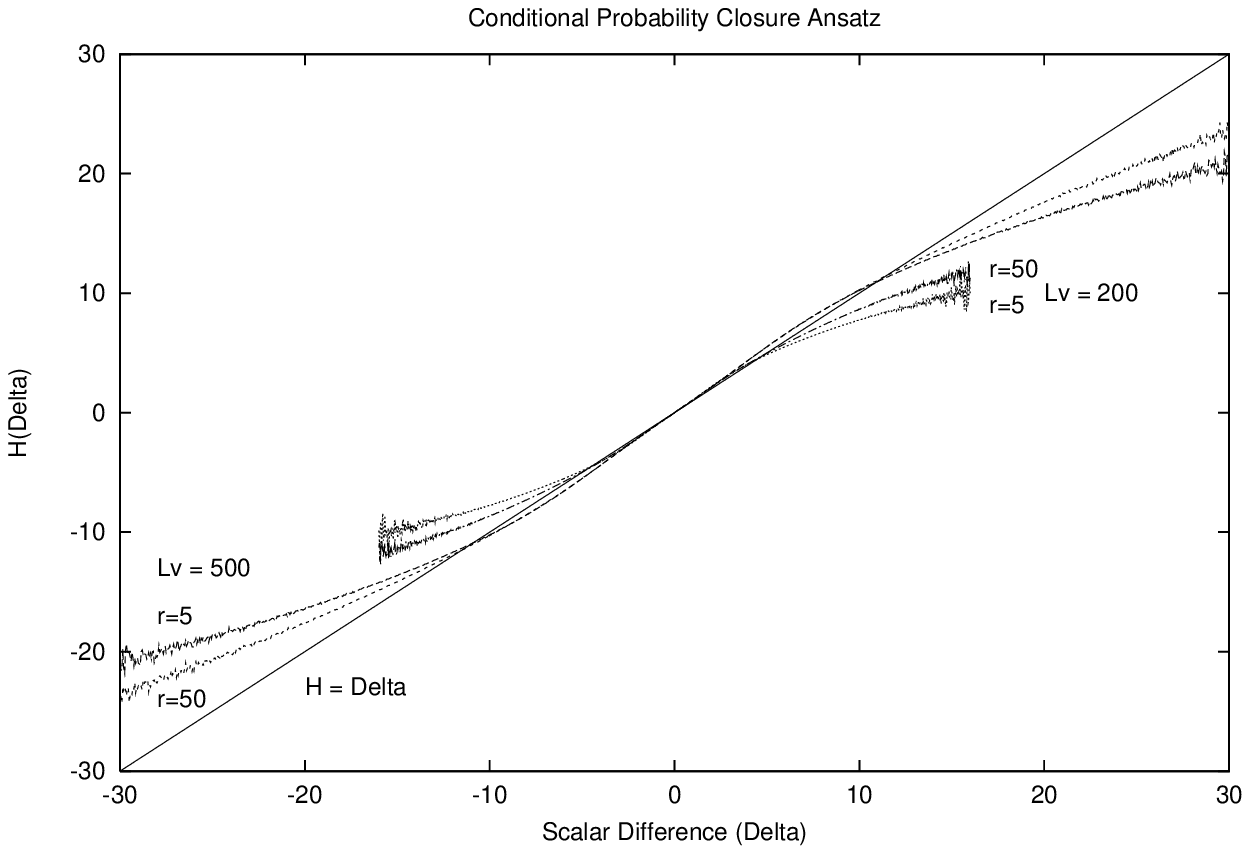}}
\small{
Conditional probability $H(\Delta,x-y) \equiv
\langle (\partial_x^2+\partial_y^2)\Delta(x,y) \vert \Delta(x,y) \rangle$ 
as a function of the temperature difference $\Delta$ for two values of $L_v$ and
two separations $r=x-y$. Data from the $y=2.1$ simulations.
$H$ is normalized so that $H=\Delta$ is the Kraichnan ansatz.
The ansatz is in good agreement with the data for small
values of $\Delta$, but the data deviates for $|\Delta| > 2gL_v$ ($g=0.01$).
This is approximately the largest value of $\Delta$ which can be generated by
a single large eddy; larger values are exponentially unlikely (see Figure \ref{fig8}).}
\caption{Conditional Probability Closure Ansatz}
\bigskip
\label{fig14}
\end{figure}

In \cite{kra95}, the closure ansatz in \cite{kra94} is derived from an assumption 
about a conditional probability, namely that
\begin{equation}
	H(\Delta,x-y) \equiv \langle (\partial_x^2+\partial_y^2)\Delta(x,y) 
		\vert \Delta(x,y) \rangle  
		= \left( {J_2 \over {2D\tau S_2}} \right) \Delta .
\end{equation} 
This conditional probability $H(\Delta,r)$ has be computed numerically in the 
simulations.  Figure \ref{fig14} shows $H(\Delta)$ as a function of $\Delta$,
normalized by ${J_2 \over {2D\tau S_2}}$,  for $\rho_2=0.1$ and both $L_v=200$ and
$L_v=500$.  Two values of $r$ are shown for each simulation.   The resulting averages
lie very close to $H(\Delta)=\Delta$ for small $\Delta$, as assumed in \cite{kra95}. 
However, they deviate from the straight line at $|\Delta| \cong 2gL_v$.  This is
approximately the point at which correlated motion becomes unimportant and the dynamics
are controlled by the `random walk' described on the previous section. The failure 
of Kraichnan's ansatz appears to be due to the existence of a finite upper size of 
the inertial range, $L_v$, which is much smaller than the system size. Other 
values of $\rho_2$ exhibit a similar behavior.

Upon closer examination, the large $n$ behavior of the constants of proportionality
$C_n$ can be related to the failure of the closure ansatz. Using the definition
of the constants, 
\begin{equation}
	C_n \equiv {J_{2n}S_2 \over nJ_2S_{2n}}
\end{equation}
the $C_n$ can be related to the closure ansatz through the definitions of $S_{2n}$
and $J_{2n}$.  These definitions are
\begin{equation}
	S_{2n} \equiv {\int P(\Delta) \Delta ^{2n} d\Delta}
\end{equation}
and
\begin{equation}
	J_{2n} \equiv nD\tau{\int P(\Delta) \Delta ^{2n-1} H(\Delta)  d\Delta}
\end{equation}
where $P(\Delta)$ is the probability distribution function for scalar differences.

If one defines the `error' in Kraichnan's closure as $\delta H$
\begin{equation}
	\delta H(\Delta) \equiv \Delta - {2D\tau S_2 \over J_2}H(\Delta)
\end{equation}
then the closure constants can be written as 
\begin{equation}
	C_n = {1-Q_{2n} \over 1-Q_2}
\end{equation}
where the functions $Q_{2n}$ are defined as
\begin{equation}
	Q_{2n} \equiv {{\int P(\Delta) \Delta^{2n-1} \delta H
		d\Delta} \over {\int P(\Delta) \Delta^{2n} d\Delta}}
\end{equation}
representing the contribution of the `error' $\delta H$ to $C_n$. 
 
Numerically, $\delta H$ is aproximately zero for small $|\Delta|$, but appears to 
grow linearly with $\Delta$ for $|\Delta| >2gL_v$.  Asymptotically, $H(\Delta)$ 
might grow as fast as linearly with $\Delta$ (but with a slope less than one).
An approximate large $|\Delta|$ form for $H$ consistent with the simulation
results is
\begin{equation}
	H(\Delta)  \simeq \alpha {J_2 \over 2D\tau S_2} \Delta
\end{equation}
where $0 \leq \alpha <1$.  Then $\delta H \propto (1-\alpha) \Delta$ for 
$|\Delta| >2gL_v$. Assuming this form for $\delta H$, the function $Q_2$ should 
be less than one. However, as $n \to \infty$, $Q_{2n} \to 1-\alpha$.  If the 
numerically observed trend for $H(\Delta)$ persists as $|\Delta| \to \infty$, this implies
\begin{equation}
	\lim_{n \to \infty} C_n = {\alpha \over 1-Q_2}
\end{equation}
which is a constant.  It would be consistent with the numerics to have $\alpha = 0$ 
and hence
\begin{equation}
	\lim_{n \to \infty} C_n = 0.
\end{equation}
It is also possible that $\alpha$ is non-zero and that ${\alpha \over 1-Q_2}$ 
depends on $\rho_2$.  This might explain the possibly different asymptotic values
for $C_n$ for different values of  $\rho_2$ seen in Figure \ref{fig13}.

Unfortunately, in the Kraichnan model the truly interesting quantity is $nC_n$.
For $\alpha =0$, there are two possible assymptotic behaviors.  If $nC_n$ is
bounded at very large $n$, the scaling exponents $\rho_{2n}$ would 
approach a constant as $n \to \infty$ (as suggested for the eddy model).  If
instead $nC_n$ is unbounded (but growing no faster than $n$), then $\rho_{2n}$
would also be unbounded, but growing no faster than $\sqrt n$. The possibility 
that $nC_n \to 0$ is excluded by the Holder inequalities, since it would imply 
$\rho_{2n}$ decreasing with $n$.  If instead $\alpha > 0$, $nC_n$ must grow linearly
with $n$ and the scaling exponents ultimately grow like $\sqrt{n}$, although the
precise values may differ from the $C_n = 1$ prediction.

\subsection{Dissipative Range Closure Ansatz}

\begin{figure}[htb]
\centerline{\epsfbox{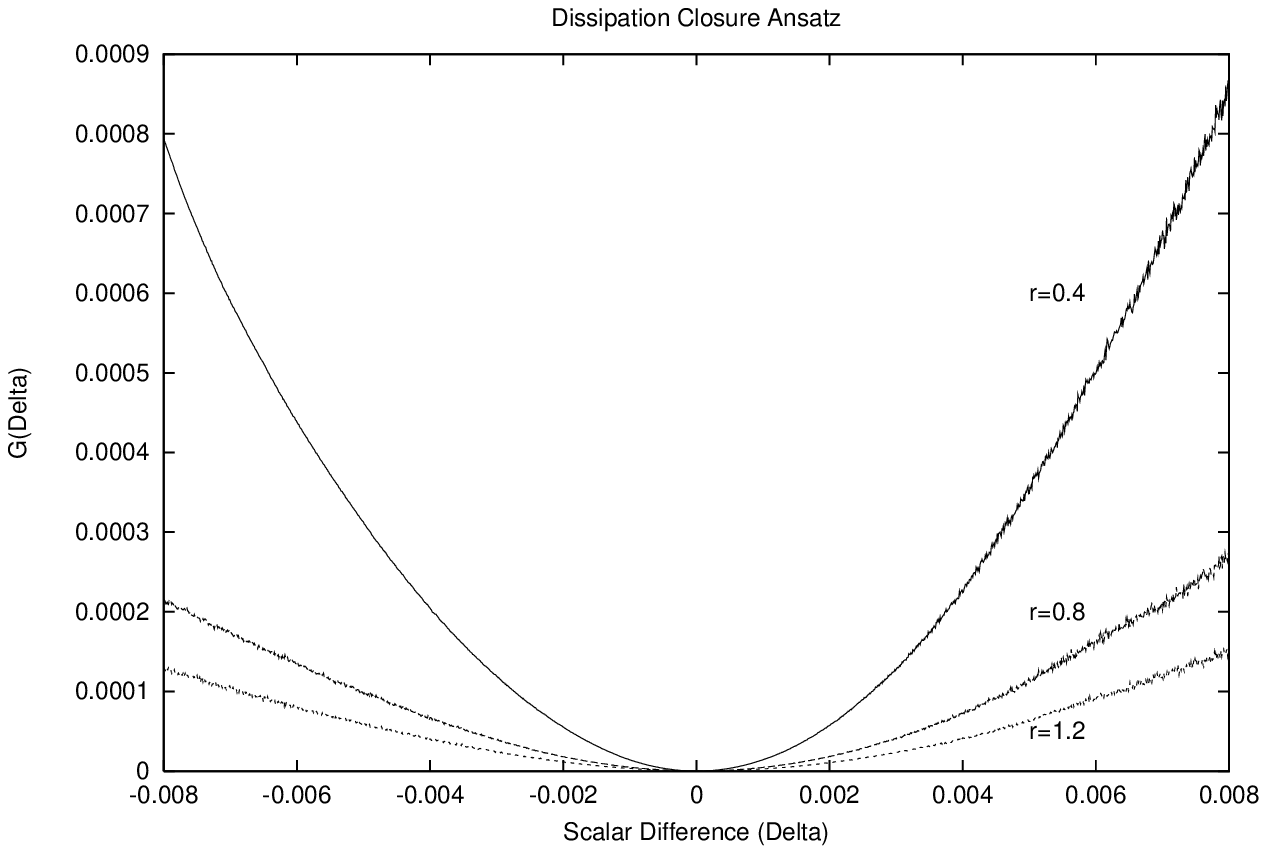}}
\small{
Conditional probability $G(\Delta,x-y) \equiv
\langle [(\partial_x\theta)^2+(\partial_y\theta)^2]|\Delta(x,y) \rangle$ 
as a function of the temperature difference $\Delta$ for several values
of $r \equiv x-y$ in the dissipative subrange.  Data from the diffusion-dominated
simulation.  The parabolic form is
consistant with the small-scale closure ansatz proposed in the previous chapter.}
\caption{Dissipation Range Closure Ansatz}
\bigskip
\label{fig15}
\end{figure}

The conditional probability $G(\Delta)$ needed for closure in the dissipative 
range is shown in Figure \ref{fig15} for several values of $r$ (from the simulation 
with $D\tau=0.02$).  The conditional probabilities are well approximated by the
parabola
\begin{equation}
	G(\Delta,r) = a + {b\Delta \over r} + {c\Delta^2 \over r^2}
\end{equation}
with $a \cong 4 \cdot 10^{-7}$, $b \cong 2 \cdot 10^{-3}$, and $c \cong 1.99$.
These values are consistent with the independently known values of $A_2$ and
$A_3$.

\begin{figure}[htb]
\centerline{\epsfbox{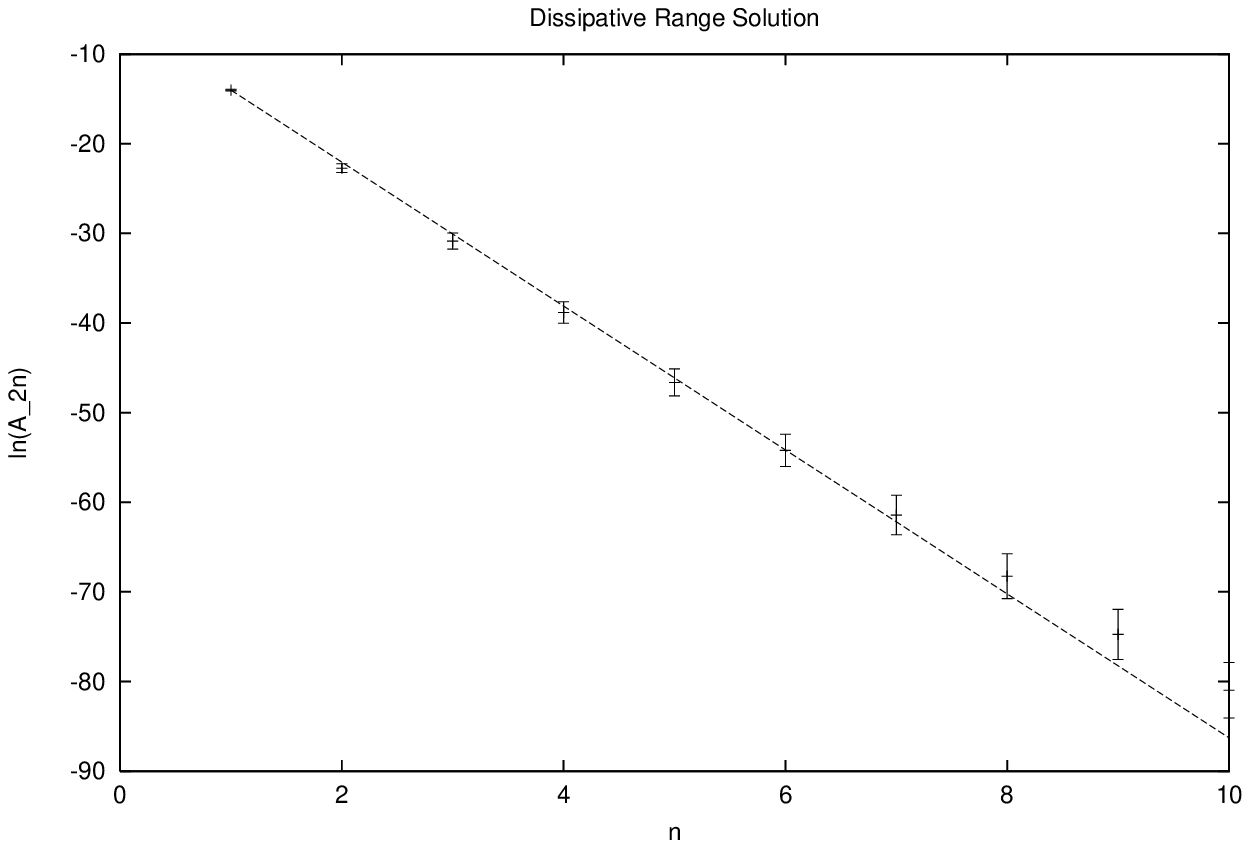}} 
\small{
Structure function coefficients $A_{2n}$ as a function of $n$
for the dissipative range regular-scaling solution, $S_{2n}=A_{2n}r^{2n}$.
The data points were determined by fitting the simulation data.
The straight line is the analytic solution, with the unknown parameter
${A_3 \over A_2}$ determined by fitting the numerical data for $S_3(r)$.}
\caption{Dissipation Range Solution}
\bigskip
\label{fig16}
\end{figure}

The structure functions are given by $S_n(r)=A_n r^n$, and the constants $A_n$ are shown in
Figure \ref{fig16}. In addition, the calculated values 
$A_n = A_{n-2}A_2 + {A_3 \over A_2}A_{n-1}$ from equation \ref{Dis}
are also shown, using the analytically known value of $A_2$ and the value of $A_3$ 
determined from fitting $S_3(r)$.  The deviations from the analytic solution at
large $n$ are possibly due to the neglected terms $F_n$ and $\cal L$ in the
equation for the structure functions.  
\clearpage

\section{Conclusions}
\subsection{Extensions to the Kraichnan Model}
The original motivation for studying this `eddy model' was to 
test the validity of the closure ansatz proposed in \cite{kra95}.
In the model, the principal failure of the closure could be traced
to the finite size of the largest eddy, $L_v$.  Specifically, the conditional 
probability $H(\Delta)$ is not a linear function of $\Delta$,
deviating at large values ($|\Delta| \gg gL_v$, where $g$ is the imposed
gradient).  The deviation arises because the largest 
velocity eddy in the system ($L_v$) is much smaller than the system size, 
so that large scalar differences only occur as a result of the uncorrelated
action of several eddies.

This finite $L_v$ effect causes the constants of proportionality $C_n$ 
proposed in \cite{chi96} to deviate
from $1$, growing with $n$ for small $\rho_2$ but shrinking with $n$ at large
$\rho_2$ (in contradiction to the predictions of \cite{chi97} and also the 
simulations of \cite{fai97}).  Other apparent consequences of this effect include
exponential tails for scalar difference PDFs and scaling exponents $\rho_{2n}$
which approach a constant value at large $n$.  
Because this effect depends only on the existence of a finite eddy size $L_v$ and 
not on the details of the smaller scale mixing, it is reasonable to expect that
it might apply generically to other models of scalar mixing, including the one
proposed by Kraichnan.
 
\subsubsection*{The limit $L_v \to \infty$}
For small values of the scalar difference $\Delta$ ($|\Delta| < gL_v$) the
numerical simulations of the eddy model appear to support the
conditional probability ansatz $H(\Delta) \propto \Delta$.  Low order structure
functions should therefore have exponents $\rho_{2n}$ which lie very close to the
values predicted by Kraichnan.  The fact that the errors become
significant at large $|\Delta|$ means that the scaling exponents $\rho_{2n}$ 
will exhibit deviations from the values predicted by this
ansatz when one looks at sufficiently high order structure functions (large $n$).
 
Mathematically, structure function scaling is defined for an infinite inertial 
range, or $L_v \to \infty$.   When considering the large $n$ behavior of the 
scaling indices, one must take the limits $L_v \to \infty$ and $n \to \infty$.
The order matters. If one considers small enough values of $n$ (for any particular
$L_v$) it will appear that Kraichnan is correct.  Hence if one lets $L_v \to \infty$
first, all values of $n$ will satisfy his predictions.  
For moderate values of $\rho_2$ the numerical simulations of the `eddy model' 
support Krachnan's ansatz if the limits are taken in this order. 
However, for any finite $L_v$ one can find sufficiently large
$n$ for which the scaling exponents deviate from the Kraichnan values
due to the finite $L_v$ effect.  Hence if one lets $n \to \infty$ first, the Kraichnan
solution will fail.  It is my personal opinion that keeping $L_v$ finite while letting 
$n \to \infty$ is the more physical limit, since any real system
will have a finite largest eddy size.

\subsubsection*{The limit $\rho_2 \to 0$}

There is some numerical evidence that regular scaling is approached in 
the limit $\rho_2 \to 0$ in the eddy model.  The constants of proportionality
deviate significantly from $C_n = 1$ even for the lowest values of $n$, 
and the lowest scaling exponents fall just below the regular scaling line
$\rho_{2n}=n\rho_2$.  In addition, the core of the scalar difference PDF
becomes very rounded, perhaps gaussian, in this limit.  The source of these
effects is not understood, but it is very possible that they have nothing 
to do with the finite $L_v$ effect.  If so, they may indicate a failure of 
the Kraichnan ansatz which would persist in the limit $L_v \to \infty$. 
However, it must be remembered that direct calculation of the conditional 
probability $H(\Delta)$ does not indicate any measurable discrepancy with
the closure ansatz for small values of $\Delta$ in this limit. 

\subsubsection*{Comparison with other Calculations}
In most of the competing calculations for the exponents of the Kraichnan model \cite
{che95} \cite{che96} \cite {gaw95}, the gradient forcing
is replaced by a random source term for the passive scalar.  In this formulation
there are two upper length scales in the problem.  They are the largest correlation
length of the velocity field, $L_v$, and the largest correlation length of the source
field, $L_s$.   Whether or not one would expect to see deviations from Kraichnan's
ansatz of the type observed in the eddy model depends on which of these two length
scales is larger.
 
First, consider $L_s \gg L_v$.  The source of the scalar is smooth over length scales
much larger than the largest motions of the velocity field.  In this case very large
scalar differences can only be generated by uncorrelated transport over scales much
larger than $L_v$.  This is analagous to the transport against the gradient by random
events in the eddy model.   Hence in this case one might expect to see scalar 
difference PDFs with exponential tails, and deviations from Kraichnan's ansatz like
those seen in the eddy model.  
  
The opposite case is  $L_s \ll L_v$.  Then the largest `globs' of scalar (of size
$L_s$) can be mixed by coherent motions of the velocity field.  The sort of
transport described above would not happen, and the Kraichnan ansatz
could be correct in this case.  However, this is the order of limits taken in
\cite{che95} \cite{che96} \cite {gaw95}, and these authors claim to find deviations 
from the Kraichnan result. If they are correct, the deviations come from a 
different source than the finite $L_v$ effect, so the eddy model
simulations don't necessarily support their claims.

A recent paper by Chertkov, {\it et. al.}, \cite{che97}, has more in 
common with the eddy model.  It considers a scalar advected by a one-dimensional
compressible velocity field with correlations analogous to the Kraichnan model. 
By assuming that the tails of the scalar difference PDF are dominated by 
the most rapidly stretched Lagrangian trajectories, they conclude that the
tails must be exponential and that the scaling exponents $\rho_{2n}$ must be
independent of $n$.  The calculations were done in the limit $\rho_2 \to 0$. 
Although the level of mathematical sophistication is quite different, 
the basic assumption is identical to the one used to understand the PDF
tails in the eddy model, and the conclusions are also similar. 

\subsection{Extensions to Pipe Flow}

As suggested earlier, one can draw physical analogies between the eddy model
and scalar mixing in turbulent pipe flow.  In particular, the separation of 
length scales (between the largest eddy size and the system size) required for
exponential tails in the scalar difference PDF can be achieved in pipe flow.  
The finite diameter of the pipe limits velocity eddies to this size, while 
the pipe itself can be much longer (analogous to this one-dimensional model).  The 
mixing in the eddy model at small scales is quite different from real flows, but
the likelihood of transport across distances larger than $L_v$ (analogous to the 
pipe diameter) could be similar (statistically) to real pipes.   This suggests
that, for large $n$, the scaling exponents in pipe flows would approach a constant
value. Further, since the slope of the PDF tails depends on the pipe diameter
($gL_v$), this value might be geometry-dependent.  

Recent experiments on pipe flows \cite{Gui97} have revealed PDF's of scalar
values with exponential tails.  No attempt has been made to study scalar 
differences, but the arguments of the previous chapter suggest that exponential
tails in the scalar value PDF would imply exponential tails in the scalar difference
PDF.  If true, this would force structure function scaling exponents	
in turbulent pipe flow to approach a constant (non-universal) value at large $n$.

Flows in other geometries (such as a box, for example), would not be expected to 
display this effect.  If the largest velocity eddy is comparable to the system size,
large-scale mixing can occur under the correlated motion of a single large eddy, and
the mechanism for generating exponential tails in the PDF would not exist.  Hence there
is no reason to expect that scaling exponents $\rho_{2n}$ would approach a constant 
at large $n$ in such geometries.  The structure function scaling exponents
of the passive scalar might therefore be geometry-dependent, with pipes exhibiting a 
different large $n$ asymptotic behavior than more open flow geometries.

\section{Acknowledgements}

It is a pleasure to acknowledge usefull discussions with L.P. Kadanoff, G. Falkovich, 
D. Lohse, P. Constantin, M. Chertkov, T. Zhou, and N. Schoegerhofer.  This work was 
supported in part by a Fannie and John Hertz Foundation Fellowship.  It also utilized 
MRSEC shared facilities under NSF-DMR grant number 9400379.

\end{document}